\documentclass{emulateapj}
\usepackage{graphicx, amsmath, amssymb, amsthm}

\makeatother

\newcommand\etal{{\it et~al.~}}
\newcommand\eg{{\it e.g.}}

\newcommand\Ls{L_\odot}
\newcommand\Ms{M_\odot}

\newcommand\Zs{Z_\odot}

\begin{document}
\title{The Molecular Hydrogen Deficit in Gamma-Ray Burst Afterglows}
\author{Daniel Whalen\altaffilmark{1}, Jason X. Prochaska\altaffilmark{2}, 
Alexander Heger \altaffilmark{3} \& Jason Tumlinson \altaffilmark{4}}
\altaffiltext{1}{Applied Physics (X-2), Los Alamos National
Laboratory; dwhalen@lanl.gov}
\altaffiltext{2}{University of California Observatories - Lick Observatory, UC Santa
Cruz, CA 95064; xavier@ucolick.edu}
\altaffiltext{3}{Theoretical Astrophysics (T-6), Los Alamos National
Laboratory}
\altaffiltext{4}{Yale Center for Astronomy \& Astrophysics, Departments of Physics \&
Astronomy, Yale University, P.O. Box 208121, New Haven, CT 06520-8121}

\begin{abstract} 

Recent analysis of five gamma-ray burst (GRB) afterglow spectra reveal the absence of
molecular hydrogen absorption lines, a surprising result in light of their large neutral 
hydrogen column densities and the detection of H$_2$ in similar, more local star-forming 
regions like 30 Doradus in the Large Magellanic Cloud (LMC).  Observational evidence further 
indicates that the bulk of the neutral hydrogen column in these sight lines lies more than 100 
pc from the GRB progenitor and that H$_2$ was likely absent prior to the burst, suggesting that 
direct flux from the star, FUV background fields, or both suppressed its formation.  We present 
one-dimensional radiation hydrodynamical models of GRB host galaxy environments, including 
self-consistent radiative transfer of both ionizing and Lyman-Werner (LW) photons, nine-species 
primordial chemistry with dust formation of H$_2$, and dust extinction of UV photons.  We find 
that a single GRB progenitor is sufficient to ionize neutral hydrogen to distances of 50 - 100 
pc but that a galactic LW background is required to dissociate molecular hydrogen in the ambient 
ISM.  Intensities of 0.1 - 100 times the Galactic mean are necessary to destroy H$_2$ in the cloud, 
depending on its density and metallicity.  The minimum radii at which neutral hydrogen will be 
found in afterglow spectra is insensitive to the mass of the progenitor or the initial mass 
function (IMF) of its cluster, if present. 

\end{abstract}


\section{Introduction}

Long-duration GRBs are extremely energetic events produced during the death of some massive 
stars.  Because massive stars have such short lifetimes ($t < 10^7$ yr), the GRB events are 
identified exclusively with star-forming galaxies \citep[$\eg$][]{ldm+03} and are expected
to occur within the star-forming regions of these galaxies.  In addition to the burst of 
gamma-ray radiation, approximately half of high $z$ GRBs emit large fluxes of rest-frame UV 
radiation observed as bright optical afterglows at Earth.  These afterglows are believed to 
result from the deceleration of a relativistic jet upon its encounter with gas at $r \sim 
10^{16}$ cm from the GRB progenitor.  The optical afterglows have peak brightnesses which 
rival (and sometimes exceed) the brightest quasars at high $z$.

In response to the discovery of optical afterglows, astronomers have initiated programs to 
obtain photometry and spectroscopy at early times ($t \lesssim 1$ hr) before these transient 
events fade below detection.  Imprinted on the power-law spectrum of a GRB afterglow are the  
absorption-line signatures of all the neutral and partially ionized gas between the GRB and 
Earth.  In principle, this includes the circumstellar medium (CSM) associated with the GRB 
progenitor, gas comprising the molecular cloud and/or \ion{H}{2} region related to the local 
star-forming region, the `ambient' interstellar medium (ISM) which surrounds the star-forming 
region, baryons in the galactic halo of the GRB host, and the so-called intergalactic medium 
(IGM) that permeates the universe.

To date, astronomers have struggled to identify signatures of the expected CSM, most likely 
because the GRB afterglow itself ionizes this gas prior to spectroscopic observations 
\citep{cpr+07,whf+07}.  A typical GRB afterglow that is bright enough for follow-up
spectroscopy emits $\sim 10^{60}$ ionizing photons during the first few minutes after onset 
(assuming isotropic emission).  This implies $\sim 10^{20}$ photons cm$^{-2}$ at 10\,pc, and 
previous calculations have shown the GRB ionizes and dissociates the majority of H and H$_2$ 
to this radius \citep{wd00,pl02,draine02}.  At distances greater than 10\,pc, however, the 
GRB afterglow should have minimal effect on the ionization state of the star-forming region 
and surrounding ISM.

The gas observed within the host galaxy of GRBs is marked by large neutral hydrogen column 
density, $n_{\mathrm{HI}}$ \citep{jfl+06}.  Many of the sightlines show $N(\mathrm{H_2}) >$
$10^{22} \, \mathrm{cm}^{-2}$, surface densities comparable to those measured for molecular 
clouds.  A recent survey for H$_2$ along a subset of these sightlines, however, has demonstrated 
that the molecular fraction $f(\mathrm{H}_2) \equiv 2 N(\mathrm{H}_2)/[2 N(\mathrm{H}_2) + 
N(\mathrm{HI})]$ is generally very low in the gas surrounding GRBs, $\lesssim 10^{-6.5}$ 
\citep{tet07} (hereafter TET07).  The authors draw two conclusions based on this observational 
result: (i) the molecular gas associated with the GRB star-forming region was photodissociated 
prior to the GRB event; (ii) the formation of H$_2$ in the ambient ISM is surpressed by an 
enhanced far-UV radiation field.  However, we note a tentative detection of H$_2$ has been 
made toward GRB 060206 with $f(\mathrm{H}_2) \sim -3.5$ \citep{fet06}.  TET07 regarded this 
result as an upper limit because they consider the identifications of the two possible H$_2$ 
absorption lines to be ambiguous.  Even if we allow this tentative discovery, the absence of
molecular hydrogen in other GRB afterglows, to very low detection limits, requires explanation.  

There are several lines of argument which suggest the gas observed in GRB sight lines 
arises primarily in the ambient ISM of the host galaxy.  First, many afterglow spectra 
exhibit strong \ion{Mg}{1} absorption at velocities identical to the (dominant) low-ion 
transitions of \ion{S}{2}, \ion{Zn}{2}, \ion{Mg}{2}, etc.  Although the GRB afterglow 
will photoionize hydrogen and these low-ion species to only $\sim 10$\,pc from the GRB,
\citet{pcb06} argued that atomic Mg should be ionized to beyond 50\,pc 
\citep[see also][]{mhk+02}.  The observation of strong \ion{Mg}{1} absorption, therefore, 
places the majority of gas at distances that probably lie beyond the local star-forming 
region.  Second, the gas along GRB sight lines also tends to exhibit transitions from 
fine-structure levels of Si$^+$, O$^0$, and Fe$^+$.  \citep{pcb06} demonstrated that this 
gas is excited by UV photons of the GRB afterglow, which sets an upper limit on its distance 
from the event of $\sim 1$\,kpc.  Third, UV pumping lends to line variability as the afterglow 
fades which allows a more precise estimation of the distance of the gas from the afterglow.  
Analysis of GRB~020813 and GRB~060418 indicates distances of $\sim 100$\,pc and $1.7$\,kpc 
respectively \citep{dcp+06,vls+07}.  Finally, as pointed out by \citet{whf+07}, independent
x-ray and optical evidence also suggests that the neutral columns are at some distance from 
the GRB but local to the source galaxy.  To summarize, the observation and analysis of 
\ion{Mg}{1} and fine-structure absorption indicates the majority of gas observed in the 
afterglow spectrum is located at $\sim 100$\,pc to 1\,kpc.

Taken together, these results inspire the following questions:  first, why is there so little 
molecular or even neutral gas within $\sim 100$\,pc of the GRB?  Does this imply an extensive 
\ion{H}{2} and photodissociation region (PDR), and if so what is the source of this radiation 
field?  If a substantial far-UV flux is suppressing H$_2$ formation in the ambient ISM prior 
to the burst, is the star-formation region associated with the GRB responsible or were other 
sources required?

We present numerical simulations of the photoionization and photodissociation of a molecular
cloud by a massive star cluster hosting a GRB progenitor.  We survey the typical radii of the
H II region and central PDR for reasonable collections of stars and evaluate the role of the
cluster and galactic UV background in suppressing H$_2$ formation in the outer regions of the
cloud.  Our paper is organized as follows: in $\S$\,2 we describe the numerical algorithm applied 
to the GRB progenitor H II region and PDR.  The evolution of the central H II region and H$_2$
abundances in the outer cloud are examined for a fiducial case in $\S$\,3.  The photoionization
of the molecular cloud core is computed for a variety of central densities and fluxes in $\S$\,4,
and the minimum LW backgrounds required to destroy H$_2$ in the cloud are tabulated in $\S$\,5.
We summarize in $\S$\,6.

\section{Numerical Algorithm}

We evolve the molecular cloud hosting the GRB with ZEUS-MP \citep{wn06, wn07a}, a massively 
parallel Eulerian reactive flow hydrocode with self-consistent multifrequency photon conserving 
UV radiative transfer \citep{wn07b} and nine-species primordial chemistry.  ZEUS-MP can transport 
photons from a point source centered in a spherical grid or plane waves along the x- or z-axes of 
cartesian or cylindrical boxes.  Updates to the gas energy equation include cooling due to electron 
collisional excitation and ionization of H, recombinations, bremmstrahlung, inverse Compton 
scattering and ro-vibrational emission by molecular hydrogen \citep{gp98}.  We also incorporate metal 
line cooling \citep{dm72} scaled to the metallicity of the gas.  

Our reaction network is the 9-species primordial gas phase model of \citet{anet97} and \citet{aet97}.  
We augmented the gas phase chemistry with dust catalysis of H$_2$ according to the prescription of 
\citet{j74}
\begin{equation}
\frac{dn_{\mathrm{H}_2}}{dt} \; = \; Z \, R\, n_{\mathrm{H}} n_{\mathrm{H}}, \label{fig:dustH2}
\end{equation}
where $R$ is estimated empirically to be $\sim$ 1 - 3 $\times$ 10$^{-17}$ cm$^{-3}$ in 
the solar neigborhood and $Z$ is the metallicity in units of $\Zs$.  In doing so, 
we make the tacit assumption that grain formation of H$_2$ is independent of temperature.  
Although this approximation is widely applied in PDR models (and is reasonable for 
T$_{\mathrm{dust}} \lesssim$ 50 K), it should be recognized that R probably falls with rising dust 
temperatures because the 'sticking factor' of H atoms on grains decreases \citep{hb79,cz02}.  
The temperature dependence of R remains poorly constrained because the grain structure is 
unknown.  This is not an issue within H II regions since dust there will be photoevaporated; 
however, because the rest of the cloud could be heated above 100 K by a galactic background, 
our results for H$_2$ production there should be considered an upper limit.

\subsection{Photon-Conserving Dust Extinction}

We modified our multifrequency UV transport to accomodate dust attenuation using the 
extinction curves of \citet{pei92} in a manner that preserves photon conservation.  In 
photon-conserving radiative transfer, the total number of absorptions in a grid zone per 
second is the number of photons entering the zone per second minus those exiting: 
\begin{equation}
N_{\mathrm{abs}} \propto 1 \: - \: e^{- \tau}, \label{eq:nabs} \vspace{0.075in} 
\end{equation}
where \vspace{0.075in}
\begin{equation}
\tau = \sum_{i=1}^{l} \sigma_i n_i \Delta r \label{eq:tau} \vspace{0.075in} 
\end{equation}
is the total optical depth of the cell (the sum is over all removal processes).  In eq 
\ref{eq:tau}, $n_i$ is the number density of the species absorbing the photon in the 
$i^{th}$ reaction, $\sigma_i$ is the interaction cross section of the reaction, and $\Delta r$ 
is the zone length in the direction of propagation. We extract the number of 
absorptions $N_{i}$ due to a particular reaction from $N_{\mathrm{abs}}$ by  \vspace{0.075in}
\begin{equation}
N_{i} \; = \; N_{\mathrm{abs}} \frac{1 - e^{-\tau_i}}{\sum_{l} (1 - e^{-\tau_l})}, \label{eq:ki} \vspace{0.075in}
\end{equation}
where the $\tau_l$ are the optical depths to all the processes and $\tau_i$ is that 
associated with the reaction.  

We incorporate dust absorptions into this procedure by adding $\tau_{\mathrm{D}}$, the local optical 
depth due to dust in the cell, to the sum in eq \ref{eq:tau}, evaluating $N_{\mathrm{abs}}$ with
eq \ref{eq:nabs}, and adding 1 - $\tau_{\mathrm{D}}$ to the sum in the denominator of eq \ref{eq:ki}:
\vspace{0.075in}
\begin{equation}
N_{i} \; = \; N_{\mathrm{abs}} \frac{1 - e^{-\tau_i}}{(1 - e^{-\tau_{\mathrm{D}}}) + \sum_{l} (1 - e^{-\tau_l})}.  \vspace{0.075in}
\end{equation}
The effect of dust absorptions is to reduce the number of all other absorptions in the 
mesh zone.  The radiative rate coefficients used in the reaction network are then easily 
derived from the $N_i$: 
\begin{equation}
k_{i} \; = \; \frac{N_{i}}{n_{i}\,V_{\mathrm{cell}}}.
\end{equation}

\subsection{Extinction Curves}

We obtain $\tau_{\mathrm{D}}$ for a mesh zone from \citet{pei92} as follows:
\begin{equation}
\tau_{\mathrm{D}}(\lambda) \; = \; \tau_{\mathrm{B}} \, \xi(\lambda), \vspace{0.075in}
\end{equation}
where $\tau_{\mathrm{B}}$ is the extinction optical depth in the $B$ band and $\tau_{\mathrm{D}}$ 
includes both absorption and scattering.  Dust and gas are not selectively transported in our model 
so $\tau_{\mathrm{B}}$ can be defined in terms of an observationally derived dust-to-gas ratio $k$  
\vspace{0.075in}
\begin{equation}
\tau_{\mathrm{B}} \; = \; k \; \times \; 10^{-21} N_{\mathrm{H}},  \vspace{0.075in}
\end{equation}
where \vspace{0.075in}
\begin{equation}
N_{\mathrm{H}} \; = \;  N_{\mathrm{HI}} + N_{\mathrm{HII}} + 2 N_{\mathrm{H_2}}, \vspace{0.075in}
\end{equation}
is the total column density of hydrogen.  The scale factor $\xi$ is given by  \vspace{0.075in}
\begin{equation}
\xi(\lambda) \; = \; \sum_{i=1}^{6} \frac{a_i}{(\lambda/\lambda_i)^{n_i} + (\lambda_i/\lambda)^{n_i} + b_i}.  \vspace{0.075in}
\end{equation}
We adopt the dust-to-gas ratio from Table 2 and parameters $a_i$, $b_i$, $n_i$, and $\lambda_i$ from 
Table 4 of \citet{pei92} for the Small Magellanic Cloud (SMC) for our opacity model, appropriate for 
the metallicities that characterize the host galaxies of GRBs \citep[$\eg$][]{pcd+07}.  We initialize 
free electron fractions in the simulations by assuming each of the metal species in the Dalgarno \& 
McCray curves is singly-ionized by a background UV radiation field. 

\subsection{Multifrequency UV Transfer/Uniform LW Background} \label{sec:fuv}

We partition photon emission rates by energy bins according to the blackbody spectra of the stars 
in the cluster.  Good convergence in frequency is found with 40 uniformly-spaced bins in energy from 
0.755 eV to 13.6 eV and 80 logarithmically spaced bins between 13.6 eV and 90 eV.  Flux in the lower 
energy range cannot ionize H or He but does drive a host of chemical reactions that regulate H$^-$ and 
H$_2$ formation and are listed in Table 1 of \citet{wn07b}.  H$_2$ photodissociation rates are 
computed with the self-shielding functions of \citet{db96} modified for thermal doppler broadening 
\citep{wn07c} to address gas flows in an approximate manner.  Multiple stars can be stationed at the 
origin but they are not spatially resolved and must all be of the same mass (and blackbody spectrum).  
Either constant luminosities or time-dependent fluxes derived from stellar evolution models can be 
used.  Stellar spectra can be updated on Courant times without serious loss of efficiency even though 
they usually evolve on much longer timescales. 

We approximate galactic FUV dissociating backgrounds with uniform fluxes that permeate the cloud, 
which are specified in multiples of $J_{21}$ $=$ 1 $\times$ 10$^{-21}$ erg cm$^{-2}$ Hz$^{-1}$ 
str$^{-1}$ s$^{-1}$ \citep{met01}.  They are implemented by a simple addition to the photodissociation 
rate coefficient in the reaction network \vspace{0.075in}
\begin{equation}
k_{\mathrm{dis}} = 1.13 \times 10^{8} \: 4 \pi J_{21}. \vspace{0.075in}
\end{equation}
The Milky Way FUV background intensity (2 $\times$ 10$^{-8}$ photons cm$^{-2}$ s$^{-1}$ Hz$^{-1}$, 
centered at $h\nu$ $=$ 12.87 eV) is related to $J_{21}$ by $I_{\mathrm{MW}}$ $=$ 32.82 $J_{21}$.

\section{H II Region/PDR Dynamics}

\begin{figure*}
\epsscale{1.17}
\plottwo{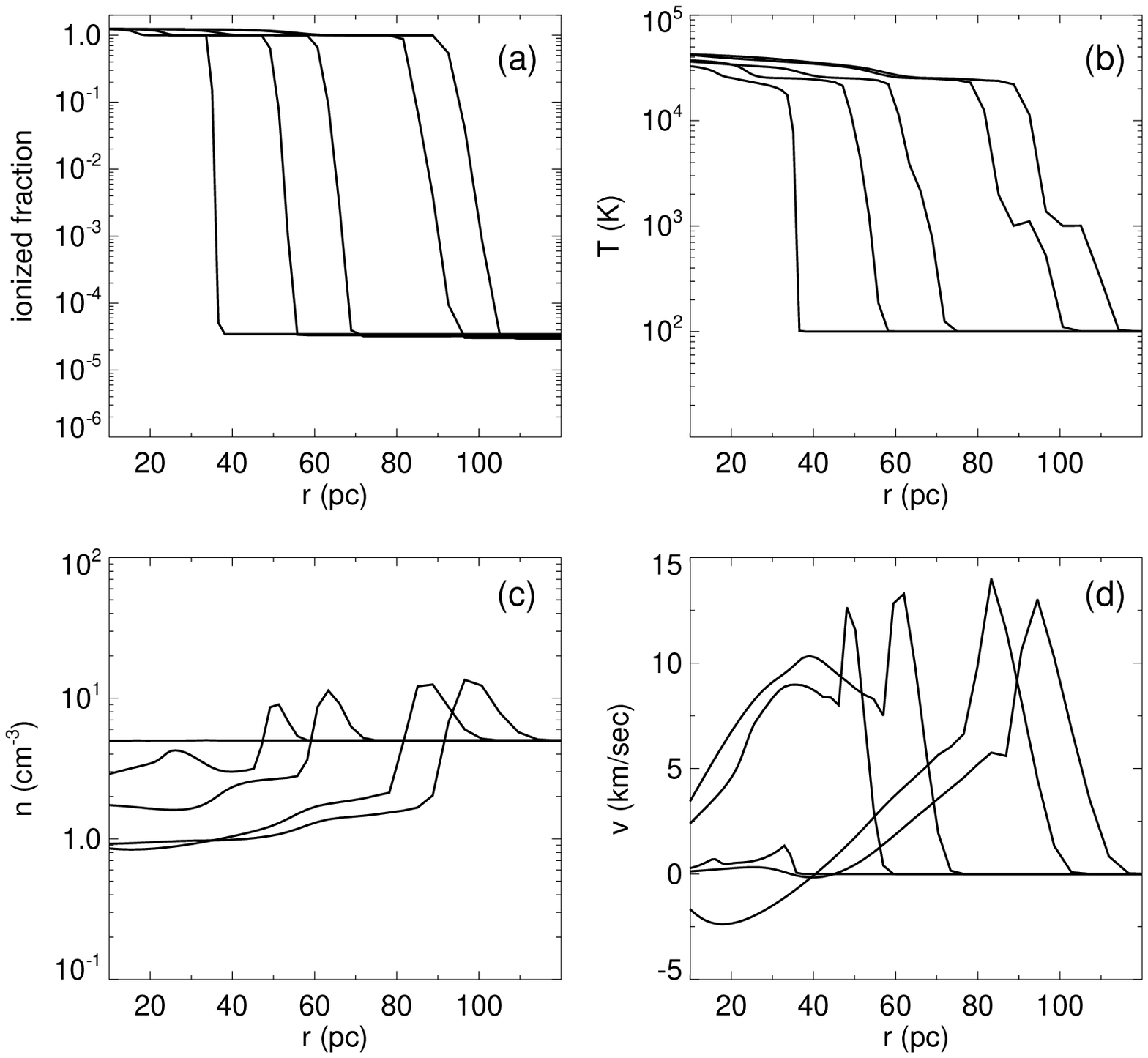}{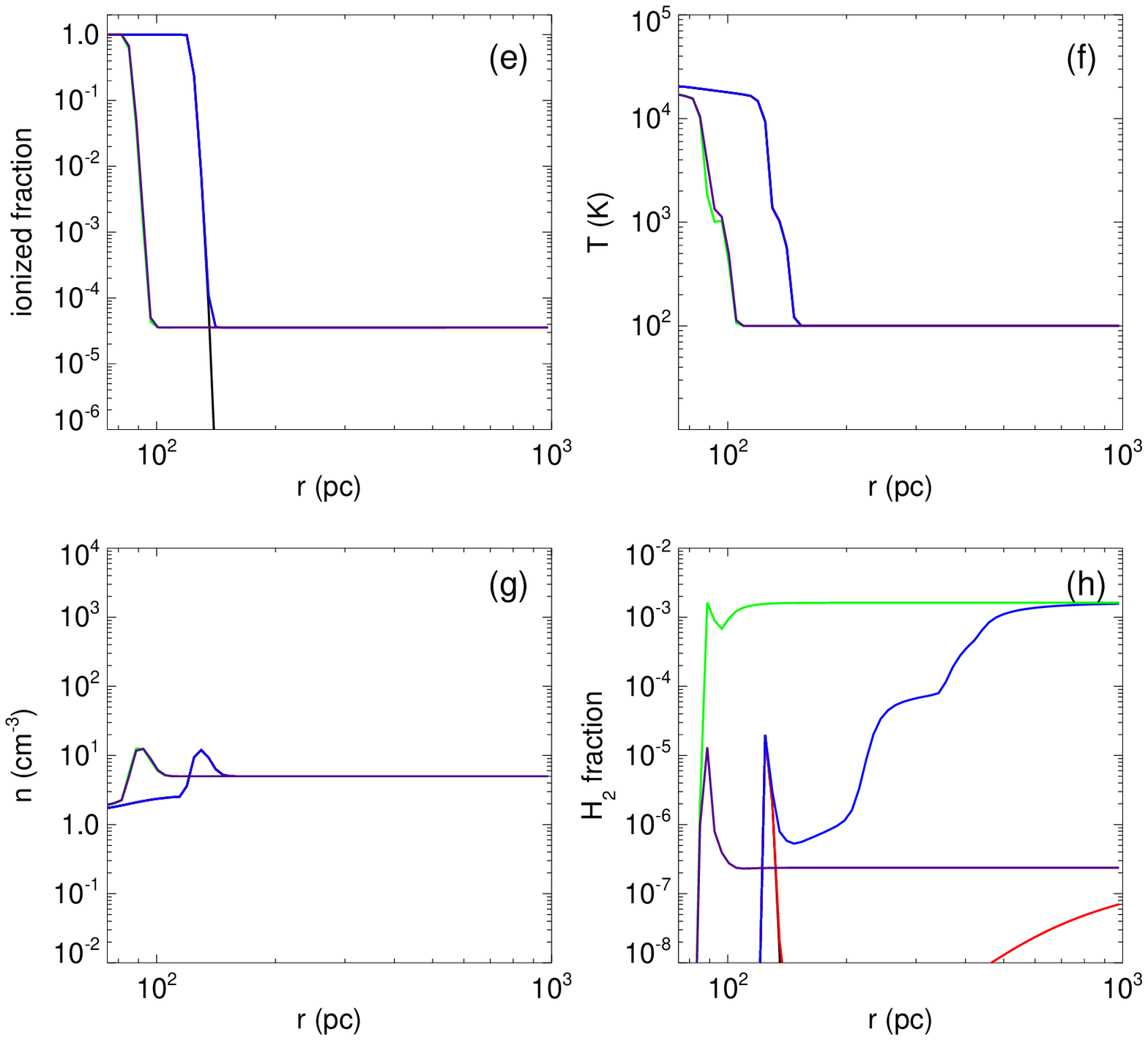}
\caption{Panels (a) - (d): density, temperature, ionization fraction, and velocity profiles of the 
photoionized GRB host cloud.  From left to right in each panel, the curves are at $t =$ 38.7 kyr, 
773 kyr, 1.55 Myr, 3.1 Myr, and 3.87 Myr, respectively.  Panels (e) - (h):  final ionized fraction, 
density, temperature, and H$_2$ fraction profiles at 3.87 Myr for the five physics test models. Black: 
model 1; blue: model 2; red: model 3; green: model 4; magenta: model 5.  In panels (e) - (g), models
1 - 3 largely overlap at the larger radii while models 4 and 5 (in which dust extinction is included)
overlap at smaller radii.  In panel (h), H$_2$ fractions are below the scale of the plot.
\label{fig:hyprof}}  
\vspace{0.33in}
\end{figure*}

We examine first the breakout of UV radiation from a cluster of massive stars in which a GRB progenitor
resides.  The stars are situated in a molecular cloud core of density $n_{\mathrm{c}}$ surrounded by a 
uniform envelope density $n_{\mathrm{cl}}$.  The density profile of the core is represented by a radial 
power law: 
\[ n(r) = \left\{ \begin{array}{ll}
			   n_{\mathrm{c}}                 & \mbox{$r \leq r_{\mathrm{c}}$} \\
			   n_{\mathrm{c}}(r/r_{\mathrm{c}})^{-\omega}   & \mbox{$r \geq r_{\mathrm{c}}$,}
                          \end{array}
                  \right.\vspace{0.1in} \]

In all the models in this paper we set r$_c$ = 0.2 pc and $\omega =$ 2, consistent with observational 
surveys of cloud cores \citep{ag85,get88}.  We smoothly join this profile to the uniform density $n_{\mathrm{cl}}$ 
assumed for the rest of the cloud at the radius at which they become equal.  This second radius varies 
with $n_{\mathrm{c}}$ and $n_{\mathrm{cl}}$ and constitutes the outer boundary of the core, since it is 
the distance at which the core becomes indistinguishable from the cloud.  The size of the cloud itself 
is just the outer boundary of the grid, as discussed below.  Here, $n$, $n_{\mathrm{c}}$, and $n_{\mathrm{cl}}$
are \textit{total} particle densities, partitioned by the mass fractions of the species assumed initially
to be on the grid, 0.76 H, 0.24 He in the models of this section.  We set this gas in hydrostatic equilibrium 
by computing the gravitational
potential whose forces exactly balance pressure forces everywhere on the grid.  Gravitational force updates 
computed from this potential are applied to the gas velocities at every hydrodynamical time step in a run 
but the potential itself is held fixed.  The self-gravity of the gas is not included in our models because 
it is negligible in comparison to the pressure forces in the ionized gas.

A source of ionizing flux was placed at the center of a one-dimensional spherical mesh, whose inner and 
outer boundaries were 0.1 pc and 1000 pc.  Reflecting and outflow conditions were assigned to the inner 
and outer boundaries, respectively.  The grid was discretized into 200 logarithmically-spaced bins 
defined by \vspace{0.1in}
\begin{equation}
\displaystyle\frac{\Delta r_{i+1}}{\Delta r_{i}} = \beta \hspace{0.25in} r_{n} - r_{1} = r_{\mathrm{outer}} - r_{\mathrm{inner}}, 
\vspace{0.1in}
\end{equation}
where n is the number of radial zones.  We applied a grid ratio $\beta$ = 1.043, which resolves the flat 
central core with 10 zones.   

\subsection{The H II Region} \label{sec:hii region}

Ten 40 $\Ms$ stars were centered in a molecular cloud with n$_c =$ 100 cm$^{-3}$ and $n_{\mathrm{cl}}$ = 
5 cm$^{-3}$, with a neutral H column of $\sim$ 1.5 $\times$ 10$^{22}$ cm$^{2}$, consistent with those of 
the GRB sight lines.  The ionizing flux of the cluster will reduce this neutral column but never by more 
than 15\,\% because the H II region rarely grows beyond 150 pc, as we will show below. The temperature 
and initial H$_2$ fraction of the gas is 100 K and 2 $\times$ 10$^{-4}$, respectively, and the UV source 
is maintained for the 3.87 Myr lifetime of a 40 $\Ms$ star.  For simplicity, we assume a uniform ionizing 
luminosity, effective surface temperature, and total luminosity of 2.469 $\times$ 10$^{49}$ photons s$^{-1}$, 
7.94 $\times$ 10$^{4}$ K, and 2.63 $\times$ 10$^{5}$ $\Ls$, respectively, for each star according to the 
zero-metallicity models of \citet{s02}.  We include a LW background equal to that of the Milky Way, a 
metallicity of 0.1 $\Zs$, and both production of H$_2$ and extinction of photons by dust.  

The ionizing UV photons of the cluster initially propagate into the cloud behind an ionization front (or 
I-front), whose thickness is only $\sim$ 20 photon mean free paths (mfp) in the neutral gas. The neutral 
fraction abruptly transitions from 10$^{-5}$ to 1 across this layer, which at first expands supersonically 
with respect to the ambient gas.  This type of I-front is known as an R-type front, and its passage leaves 
the gas ionized and hot, but otherwise undisturbed.  However, the I-front eventually slows down due to
geometric dilution of the photons reaching the front and recombinations in the ionized gas, or H II region.  
Recombinations to the ground state absorb a photon from the cluster but release another ionizing photon, 
with no net effect on the advance of the front.  Recombinations to excited states absorb photons from the 
cluster without returning any ionizing photons to the I-front and slow its expansion.  In a static medium, 
when the number of recombinations to excited states in the entire H II region equals the number exiting the 
cluster, the I-front has reached the Str\"{o}mgren radius and can advance no further.

In a real gas, a pressure wave builds in the H II region.  As the pressure wave accelerates and the front 
decelerates, the wave overtakes the front at approximately the Str\"{o}mgren radius, steepens into a shock, 
and breaks through the front.  The pressure of the ionized gas drives the neutral shock out into the cloud.  
As the shock grows in radius, it sweeps up gas in the cloud, which accumulates in a dense shell that detaches 
the shock from the I-front.  This kind of I-front is known as a D-type front, and it advances supersonically 
with respect to the cloud, but subsonically with respect to its own ionized gas.

A good approximation for the initial Str\"{o}mgren radius of the H II region (which accounts only for hydrogen
ionizations) is
\begin{equation}
r_{\mathrm{S}} = \left(\displaystyle\frac{3 {\dot{n}}_{\mathrm{ph}}}{4 \pi {\alpha}_{\mathrm{B}} {{n_{\mathrm{H}}}^2}} \right)^{1/3},
\end{equation}
where ${\dot{n}}_{\mathrm{ph}}$ is the emission rate of ionizing photons in s$^{-1}$, ${\alpha}_{\mathrm{B}}$ 
is the case B recombination coefficient of H, and $n_\mathrm{H}$ is the number density of H.  The final 
radius of the H II region (after coming to pressure equilibrium with its surroundings) is
\begin{equation}
r_{\mathrm{f}} \; = \; r_{\mathrm{S}} \left(\frac{2 T_{i}}{T_0}\right)^{2/3},
\end{equation}
where T$_i$ and T$_0$ are the ionized gas and ambient temperatures, respectively.  Given our model 
parameters, r$_S$ $\sim$ 12 - 90 pc, depending on whether core or cloud densities are used, and r$_f$ 
is $\sim$ 85 times greater.  Thus, the H II region never comes into pressure equilibrium in the life 
of the association and is still expanding when the GRB explodes.  As we later show, dust extinction 
reduces r$_s$ by a third in our runs.  We adopt a cooling cutoff of 1000 K, defined to be the gas 
temperature below which cooling curves are not applied to the gas energy updates.  In reality, the 
Dalgarno \& McCray curves would reduce the temperature of the shocked shell driven by the ionization
front (I-front) to below 100 K, collapsing it to higher densities and narrower widths than in our 
models.  However, this results in less forgiving Courant times and alters neither the escape of UV 
nor the dynamics of the H II region, so we proceed with the higher cutoff.  

Density, temperature, ionization fraction, and velocity profiles are shown in Figs \ref{fig:hyprof}a-d 
at $t =$ 38.7 kyr, 773 kyr, 1.55 Myr, 3.1 Myr, and 3.87 Myr.  The I-front at first propagates so rapidly
that no hydrodynamical response of the gas can overtake it.  This is an R-type front, and its supersonic
nature is evident at 38.7 kyr:  the gas is barely disturbed and the growing velocity wave lags the front. 
As the I-front expands it slows down due to both geometrical dilution of radiation and recombinations in
the ionized gas that reduce the number of ionizing photons reaching the front.  The pressure wave building 
up in the hot interior catches up to the front, pushes past it, and steepens into a strong shock.  The 
I-front is now D-type, and as it continues to grow it sweeps up shocked neutral gas in a dense shell that 
detaches the shock from the front.  In this model, the I-front executes a transition from R-type to D-type 
at 460 kyr.  The relatively large postfront gas temperatures of $\sim$ 35,000 K are due to the ionization 
of He by the hard UV spectrum of the star cluster.  In the large densities of the core, the spread in mean 
free paths of the UV photons is fairly narrow so the I-front is barely resolved.  As it reaches the lower 
intracloud densities the front broadens to approximately 5 pc. After becoming D-type, the I-front drives 
the shock into the neutral medium at 13 km s$^{-1}$.  As the H II region expands, its densities fall 
from 100 cm$^{-3}$ to less than 1 cm$^{-3}$, so the GRB explodes in a comparatively diffuse environment.  
By 3.1 Myr the shock separates from the I-front, as shown in the temperature profile.  The densities in 
the neutral shell snowplowed by the front rise to $10$ cm$^{-3}$.

At the cooling cutoff temperature the shocked shell remains fairly thick, approximately
20 pc.  Even so, in three dimensions the I-front would be prone to violent dynamical instabilities
mediated by radiative cooling in the shell by metal lines \citep{wn07a}.  The cooling fractures the 
shell, allowing radiation to jet forward through the fissures in prominent fingers of ionized gas much
larger than the original perturbations.  These instabilities would form even if no metals were present 
\citep{wn07b} because the I-front forms H$_2$ at the base of the shocked neutral shell.  As we show in 
the model 5 curve in Fig \ref{fig:hyprof}h, molecular hydrogen formed via the H$^-$ channel in the partially 
ionized outer layers of the broadened I-front is not entirely destroyed by Lyman-Werner flux from the cluster
\citep[section 2 of][]{wn07b}.  The 2 $\times$ 10$^{-5}$ H$_2$ abundances, although not as efficient as
metals, still sufficiently cool the shell for instabilities to erupt in the I-front 
\citep[see Fig 5 of][]{wn07b}.  These features are relevant to the post-afterglow evolution of the GRB 
because they form clumps that would be advected through the shock front.  However, it is unlikely that 
these clumps would harbor enough H$_2$ for detection along sight lines. 

\subsection{The Photodissociation Region} \label{sec:pdr}

We performed five runs in which elements of our physics model were sequentially adopted in order 
to evaluate their individual impact on H II region dynamics, the central PDR, and equilibrium H$_2$ 
abundances in the outer cloud.  In the first run no initial electron fraction ($\chi_e$), dust 
production of H$_2$, dust extinction of photons, or Lyman-Werner background radiation were present.
The H II region and PDR would be largest in this model because no dust absorbs photons
and there are no free electrons to restore H$_2$ lost to dissociation by the cluster.  In the second 
model, an initial $\chi_e$ consistent with singly-ionized metal species was included, so H$_2$ levels 
beyond the H II region attain an equilibrium set by photodissociation by central LW photons and 
formation by the H$^-$ and H$_2^+$ channels in the gas phase:     
\vspace{0.1in}\[
\begin{array}{ccccccc}
\mathrm{H} +  \mathrm{e}^{-} & \rightarrow & \mathrm{H}^{-} + \gamma &  & \mathrm{H}^{-} + 
\mathrm{H} & \rightarrow & \mathrm{H}_{2} + \mathrm{e}^{-} \\
     &   \\
\mathrm{H}^{+} + \mathrm{H}  & \rightarrow & \mathrm{H}_{2}^{+} + \gamma &  & \mathrm{H}_{2}^{+} + 
\mathrm{H} & \rightarrow & \mathrm{H}_{2} + \mathrm{H}^{+} 
\end{array} \vspace{0.1in} \] 
The H$^-$ channel is dominant in these density and temperature regimes and will restrict the final 
size of the PDR by replenishing H$_2$ lost to LW photons from the center.  The activation of H$_2$ 
formation by dust grains in model 3 further reduces the PDR.  Neither of these processes affect the
size of the H II region, but activation of UV absorption by dust in model 4 will slow the front. In 
model 5 we add a uniform LW flux equal to the Milky Way background that dissociates H$_2$ everywhere 
in the cloud.  It is the fiducial case discussed above.  We summarize the five models in Table 
\ref{tbl-1}.

\begin{deluxetable}{ccccc}
\tabletypesize{\scriptsize}
\tablecaption{ Five GRB Host Cloud Models\label{tbl-1}}
\tablehead{
\colhead{model} & \colhead{$\chi_e$} & \colhead{dust H$_2$} & \colhead{dust extinction} & \colhead{LW Background}}
\startdata
 1 & no  & no  & no  & no  \\
 2 & yes & no  & no  & no  \\
 3 & yes & yes & no  & no  \\
 4 & yes & yes & yes & no  \\
 5 & yes & yes & yes & yes \\
\enddata
\end{deluxetable}

\begin{figure*}
\epsscale{1.17}
\plottwo{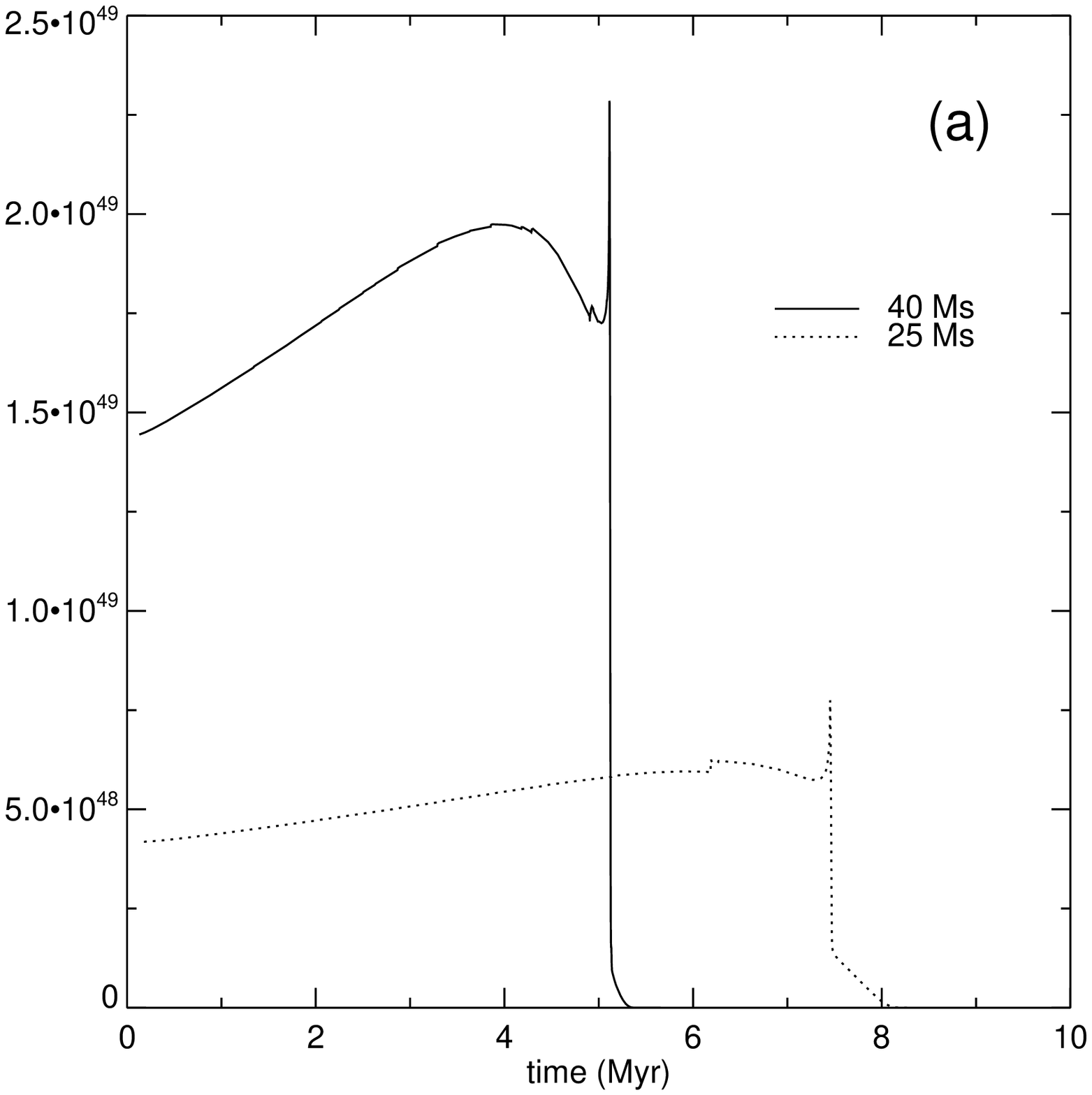}{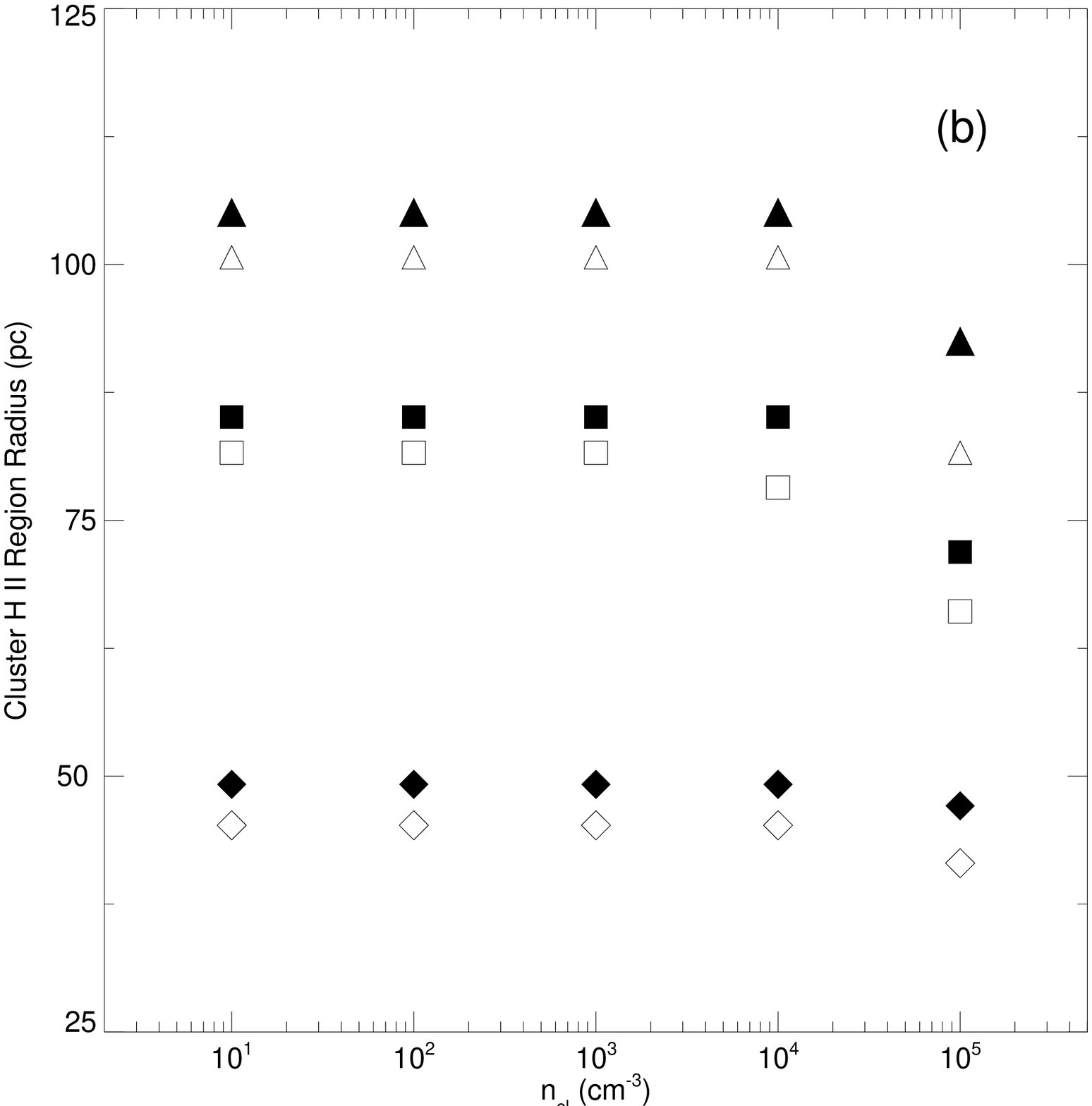}
\caption{Panel (a): emission rate of all ionizing photons as a function of time. Panel (b): radii of the 
cluster H II regions at the time of the gamma-ray burst.  Filled symbols: 25 $\Ms$ cluster; unfilled symbols: 
40 $\Ms$ cluster.  Triangles:  $n_{\mathrm{cl}}$ = 5 cm$^{-3}$; squares:  $n_{\mathrm{cl}}$ = 10 cm$^{-3}$; 
diamonds:  $n_{\mathrm{cl}}$ = 50 cm$^{-3}$.  
\label{fig:HIIregions}}  
\end{figure*}

Ionized fractions, densities, temperatures and H$_2$ abundances for the cloud at 3.87 Myr appear in 
Figs \ref{fig:hyprof}e-h.  The flow profiles clearly fall into two classes:  those that incorporate dust 
extinction and those that do not.  We find that dust attenuation at abundances similar to the SMC 
reduce the radius of the H II region by 30\,\% at the end of the main sequence lifetime (MSL) of the 40 
$\Ms$ star, from 120 pc to 85 pc.  Lower photoionization rates drop postfront gas temperatures by 
a few thousand K, which in turn slightly reduces the shock velocity.  Aside from this there are few 
differences in the structure of the flow curves of the two groups: shocked shells and I-fronts have 
nearly the same width at 3.87 Myr. 

On the other hand, each additional physical process significantly alters H$_2$ fractions in the cloud.
The dashed curve represents model 1, in which no free electrons or dust form H$_2$, and it peaks at
125 pc (being overlaid by the model 2 and 3 curves) but then drops to very low levels beyond.  The 
peak is the molecular hydrogen manufactured in the front; it cannot shield the initial 2 $\times$ 
10$^{-4}$ H$_2$ fractions in the outer region of the cloud from LW flux from the cluster, so they are
 mostly destroyed.  Molecular hydrogen within the H II region is collisionally dissociated by free 
electrons to extremely low levels.  The small free electron fraction ($\sim$ 3.5 $\times$ 10$^{-5}$) 
maintains a little of the original H$_2$ beyond the H II region at 500 pc in model 2 as shown by the 
dotted curve, but the entire cloud is still essentially dissociated by the cluster.  Activation of H$_2$ 
production by dust shields the outer cloud from LW flux, and molecular hydrogen abundances there rise by 
an order of magnitude above their initial values beyond 500 pc.  The PDR in this model has a radius of 
250 pc.

In model 4, dust attenuation of photons absorbs all central LW flux within 5 pc of the H II region, so 
H$_2$ fractions beyond rise by a factor of ten in 3.87 Myr, to 2 $\times$ 10$^{-3}$.  Molecular hydrogen 
abundances in the I-front are 100 times greater than in models 1, 2, and 3 due to FUV absorption by dust 
in the H II region.  In reality, this peak in H$_2$ would subside over time as the dust is evaporated by 
ionized gas, so the PDR would expand slightly more than 5 pc beyond the H II region.  The uniform LW 
background activated in model 5 reduces H$_2$ fractions in the I-front peak to 2 $\times$ 10$^{-5}$ and to 
2 $\times$ 10$^{-7}$ in the rest of the cloud, lowering H$_2$ mass fractions below 10$^{-6.5}$, 
the detection limit in the GRB sight lines.  All four processes are key to constructing accurate molecular 
hydrogen abundances in the GRB host cloud.  Models 4 and 5 clearly demonstrate that H$_2$ abundances in the 
cloud are set by the LW background, since for realistic models of the cluster the central PDR is confined 
to within a few pc of the H II region by dust absorption (rather than by H$_2$ formation in either the gas 
phase or on dust grains).  Our models demonstrate that the sizes of the H II regions, and therefore both the 
column density of neutral H and the minimum radii at which it is observed, are decoupled from what sets H$_2$ 
abundances in the cloud.  In the next section we examine the minimum distances from the GRB at which neutral 
H would be detected for a variety of clusters and cloud core densities.  In the section thereafter, we 
evaluate the minimum galactic UV backgrounds required to drive H$_2$ below detection limits.

\section{The GRB Host Cloud H II Region} \label{sec:hii regions}

\subsection{Clusters} \label{sec:clusters}

For what range of central densities and stellar masses is the molecular cloud core completely ionized, and 
what final H II region radii result?  We consider two clusters of ten stars, one composed of 25 $\Ms$ stars 
and the other with 40 $\Ms$ stars.  For $n_{\mathrm{cl}}$ = 5, 10, and 50 cm$^{-3}$ we consider central core 
densities of 10, 10$^2$, 10$^3$, 10$^4$, and 10$^5$ cm$^{-3}$, a total of 30 models.  The $n_{\mathrm{c}}$ 
are consistent with molecular cores observed in the Galaxy, and $n_{\mathrm{cl}}$ = 5 - 10 cm$^{-3}$ are  
typical of clouds in the Milky Way and yield the neutral hydrogen columns of the GRB spectra given our mesh
boundaries.  We include $n_{\mathrm{cl}}$ = 50 cm$^{-3}$ as a reasonable upper limit in case average cloud 
densities rise with redshift, 

For consistency with the metallicity of the GRB host galaxy, we adopt the time-dependent spectra of 0.1 $\Zs$
25 and 40 $\Ms$ stars in this parameter survey.  At a given mass, a primordial star is more compact and hotter 
than a 0.1 $\Zs$ star for two reasons.  First, Population III stars derive their nuclear energy primarily from
inefficient proton-proton burning, with only a small contribution from the CNO cycle due to small amounts of C
that are built up in the pre-main sequence phase \citep{cast83,eleid83}.  Since at a given temperature this 
process has lower energy generation rates, the core must rise to higher temperatures to support the star from 
gravitational collapse.  This, together with the fact that there are no metals to absorb photon momentum in the 
photosphere and expand and cool it, makes primordial stars considerably hotter than metal-enriched stars of
equal mass \citep{tuml00}.    
The time-dependent spectral fluxes for these stars were computed using the KEPLER code \citep{wzw78} assuming a 
time-dependent black body spectrum at the effective temperature of the star.  The result is shown in Fig 
\ref{fig:HIIregions}a.  At end of central hydrogen burning a very brief rise in the ionizing flux due to 
contraction of the star is observed, but then the stars become giants and the ionizing flux drops quickly 
to levels that are essentially negligible when compared to those when the star is on the main sequence.  We 
implement these spectra in the code as photon emission rates that are binned by energy, as described earlier.  
Notice that the total emission rates of both stars are lower than those of \citet{s02} because they are somewhat 
cooler, and that their luminosities can vary by 30\,\% over their lifetimes.  The 0.1 $\Zs$ stars also have longer 
lives:  5.65 Myr \textit{vs} 3.86 Myr for 40 $\Ms$ and 8.32 Myr \textit{vs} 6.46 Myr for 25 $\Ms$.  The coordinate 
grid and all other model parameters were those used in section \ref{sec:hii region}.

H II region radii at the time of the GRB for all 30 runs are shown in Fig \ref{fig:HIIregions}b.  Several trends 
are apparent.  First, the cloud density $n_{\mathrm{cl}}$ determines the H II region radius when $n_{\mathrm{c}}$ 
is small because the front quickly ionizes the core.  When central densities are greater than 10$^4$ cm$^{3}$, the 
front requires a significant fraction of the lifetime of the star to break out of the core so $n_{\mathrm{c}}$ 
begins to also regulate the final radius.  At any core density, the H II region shrinks with larger $n_{\mathrm{cl}}$ 
due to its greater inertia, as expected.  Interestingly, for a given $n_{\mathrm{c}}$ and $n_{\mathrm{cl}}$ the less 
massive cluster has the larger H II region.  Overall, the H II region radius at the time of the GRB varies little: 
from 41.1 to 100.7 pc over four orders of magnitude in $n_{\mathrm{c}}$, an order of magnitude in $n_{\mathrm{cl}}$, 
and a factor of five in luminosity. 

This somewhat surprising lack of variation in radius is due to two factors.  First, if the source 
spectrum can ionize helium then postfront gas temperatures are $\sim$ 25,000 K, varying little with 
the number of stars in the cluster and only somewhat with their surface temperatures.  This confines 
the sound speed $c_{\mathrm{s}}$ of the ionized gas, and therefore the expansion rate of the D-type front, 
to a fairly narrow range because of the weak dependence of $c_{\mathrm{s}}$ on T.  Since the I-front will become 
D-type in a small fraction of the life of the star in realistic clusters because of dust, the velocity 
of the front is determined by the sound speed of the gas rather than the size or luminosity of the cluster.
Dust augments the role of recombinations in the ionized gas by absorbing photons before they can reach 
the I-front and advance it, hastening its transition to D-type.

Second, the time for which the H II region has grown when it is illuminated by the afterglow is, to 
first order, the lifetime of the progenitor because stars in the cluster form at approximately the 
same time.  Since main sequence lifetimes for likely progenitors vary by no more than a factor of 
two, we expect neutral H in general to be observed 50 - 100 pc from the GRB.  We note that since the 
final radius is determined by both the life of the cluster and the sound speed of the ionized gas, 
the 25 $\Ms$ cluster has the larger H II region simply because its stars live longer.  

A simple estimate of the size of the H II region at the time of the burst follows from \vspace{0.075in}
\begin{equation}
c_{\mathrm{s}} \, = \, \left(\frac{\gamma k T}{\mu m_{\mathrm{H}}}\right)^{1/2}.  \vspace{0.075in}
\end{equation}
For completely ionized gas ($\mu =$ 1.70) at 25000 K, $c_s \sim$ 14.3 km s$^{-1}$.  With MSL ranging from 
3.87 to 6.46 Myr, the final radii of the H II regions would fall between 56 and 94.3 pc, similar to those 
found by ZEUS-MP.

\begin{figure}
\resizebox{3.45in}{!}{\includegraphics{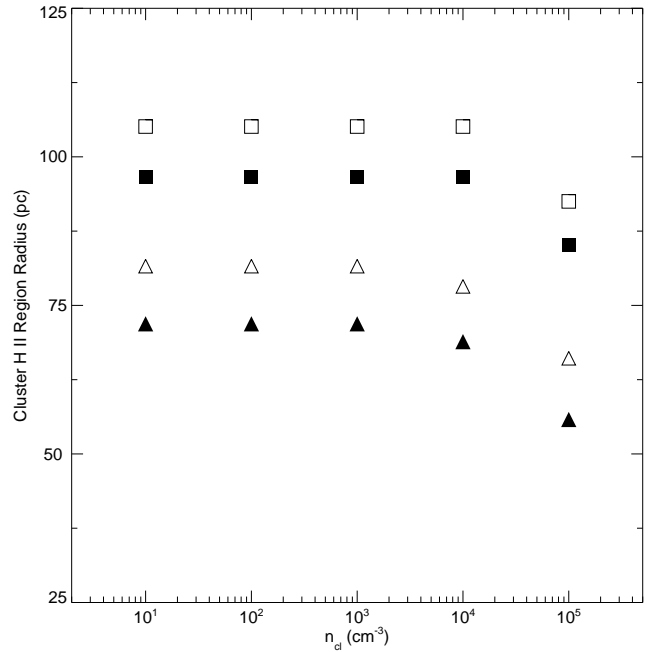}}
\caption{Radii of H II regions with constant luminosities (unfilled symbols) and time-varying spectra 
(filled symbols).  Triangles: 40 $\Ms$ stars with $n_{\mathrm{cl}}$ = 10 cm$^{-3}$; squares: 25 $\Ms$ 
stars with $n_{\mathrm{cl}}$ = 5 cm$^{-3}$.
\label{fig:timev}}  
\end{figure}

Comparison of the sizes of the H II regions in models 3 and 4 of section \ref{sec:hii region} suggests
that the H II regions of clusters in higher redshift galaxies as a rule may have been larger than those 
in the Galaxy today because there was less dust to attenuate photons.  We neglect the evaporation of dust 
within the H II region and therefore somewhat underestimate the flux that reaches the ionization front, so 
these are lower limits for their radii.  However, models 3 and 4 in section \ref{sec:hii region} show that 
this effect is at most 30\,\% (and probably less, because dust on both sides of the I-front is not immediately 
destroyed and will still retard it).  

The Milky Way LW background assumed in all 30 runs efficiently dissociates any H$_2$ in the higher densities 
between the H II region and the outer boundary of the core.  Beyond this boundary, molecular abundances 
assume the same low equilibrium values as in model 5 of section \ref{sec:pdr}, regardless of $n_{\mathrm{c}}$.  
Again, we emphasize that these are not the ultimate sizes of the H II regions because, far from being in 
pressure equilibrium with their environments, they continue to expand as long as there is sufficient flux 
to maintain ionization.  These are simply their radii at the time of illumination by the GRB.  

\begin{figure}
\resizebox{3.45in}{!}{\includegraphics{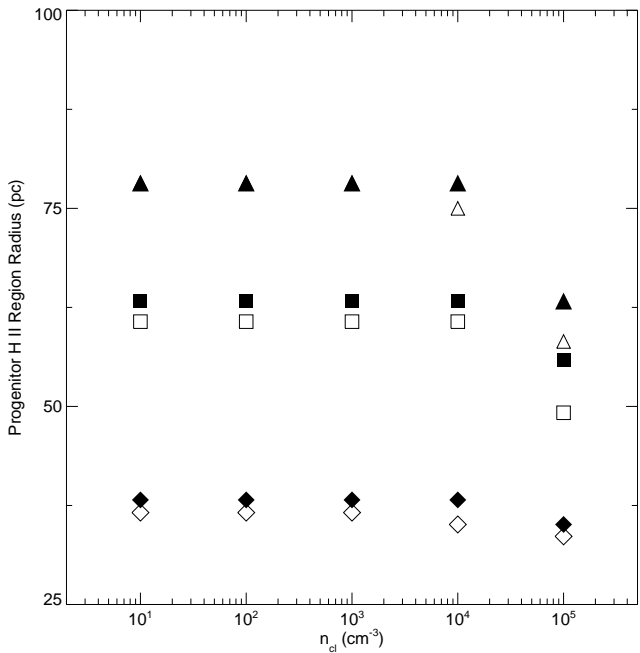}}
\caption{Radii of the progenitor H II regions at the time of the gamma-ray burst.  Filled symbols: 
25 $\Ms$ star; unfilled symbols: 40 $\Ms$ star.  Triangles:  $n_{\mathrm{cl}}$ = 5 cm$^{-3}$; squares:  
$n_{\mathrm{cl}}$ = 10 cm$^{-3}$; diamonds:  $n_{\mathrm{cl}}$ = 50 cm$^{-3}$.  Note that the sizes of
the H II regions of both progenitors match in $n_{\mathrm{cl}}$ = 5 cm$^{-3}$ for $n_{\mathrm{c}} < $ 
10$^4$ cm$^{-3}$
\label{fig:HIIregions2}}  
\end{figure}

\subsection{Stellar Spectra}

We have shown that the masses of the stars in the cluster have little effect on the final sizes of GRB
host H II regions.  Now, for completeness, we examine the H II regions of both zero-metallicity and 0.1 
$\Zs$ stellar stars.  This range in metallicity effectively brackets that of the host galaxies of the 
observed afterglows: 5.5 $\times$ 10$^{-3}$ - 0.234.  The variation in I-front evolution over these
metallicities will probably be less than that over the stellar masses chosen for this study, since 
temperatures and luminosities change less with composition than with mass.  In doing so, we also address
in an approximate way the consequences of assuming a black body radiation spectrum without correcting for
the stellar atmosphere.  Changes in the flux due to the presence of an atmosphere are less severe than 
the difference in flux between 25 and 40 $\Ms$ stars.      

In Fig \ref{fig:timev} we compare the final H II region sizes for both metal-free and 0.1 $\Zs$ 25 and 
40 $\Ms$ clusters.  We take the emission rate of ionizing photons, surface temperature, total luminosity, and 
main sequence lifetime of the 25 $\Ms$ stars to be 7.583 $\times$ 10$^{48}$ s$^{-1}$,  7.08 $\times$ 10$^{4}$ 
K, 7.76 $\times$ 10$^{4}$ $\Ls$, and 6.459 Myr, respectively \citep{s02}.  The 25 $\Ms$ clusters are in cloud
densities $n_{\mathrm{cl}}$ = 5 cm$^{-3}$ and the 40 $\Ms$ clusters are in $n_{\mathrm{cl}}$ = 10 cm$^{-3}$.  
It is clear that time variation in the luminosity alters the final sizes of the H II regions by at most 15\,\%.  
The 0.1 $\Zs$ clusters have slightly larger H II regions because of their longer lives.  Since the sizes of the 
H II regions are determined by the lifetimes of the stars and the sound speed of the ionized gas, H II regions 
of clusters with metals should be larger than those of clusters without metals by a factor equal to the ratio 
of their lifetimes.  This is greater than what we found because the spectra of the 0.1 $\Zs$ clusters soften 
as their stars cool with age, lowering $c_{\mathrm{s}}$ and slowing the expansion of the I-front.

\subsection{Solitary Progenitors}

So far, we have shown that as long as there is enough UV flux to break out of a molecular cloud core and
establish an H II region, its final size is largely independent of total flux or spectral profile. Could
then a single GRB progenitor create the H II regions implied by the minimum distances at which neutral H
is found in the afterglow spectra?  We repeat the models of section \ref{sec:hii regions} but use just one
star instead of 10.  The sizes of the H II regions are shown in Fig \ref{fig:HIIregions2}.  

Interestingly, the H II regions are similar to those of the clusters, attesting to the fact that postfront 
gas temperature rather than flux governs their sizes.  They also follow the same trends: their radii fall 
off somewhat with increasing $n_{\mathrm{c}}$ at a given cloud density and even more sharply with 
$n_{\mathrm{cl}}$.  Final radii are greater for smaller stars due to their longer lives.  The H II region 
of the 25 $\Ms$ star is slightly more sensitive to the core density, particularly when cloud densities are 
small.  We conclude that solitary progenitors are as effective as clusters in forming H II regions in the 
GRB host galaxy.

Our H II region radii are insensitive to $r_{\mathrm{c}}$ and realistic choices of density gradient $\omega$.
In section 3.1 we showed that even if H II regions expand in core densities (10$^5$ cm$^{-3}$) rather than in 
cloud densities (10 - 50 cm$^{-3}$), they reach Str\"{o}mgren radii $>$ 10 pc and final equilibrium radii 
that are many times larger.  It is clear that extending r$_{\mathrm{c}}$ to 1 pc, which is unrealistically 
large for the centers of galactic molecular cloud cores, would have very little effect on final H II region
radii.  Furthermore, the I-front evolution is very similar for density gradients of 1.5 $> \omega >$ 2.5 
\citep{ftb90}, so there will be little variation in final H II region radii over reasonable ranges of core
density profiles.

\subsection{Dust Extinction}

The simulations described so far all assume SMC dust extinction for UV transport.  How much smaller 
would the H II regions be if more dust is present in the GRB host galaxy?  We calculated the H II 
region of a single 25 $\Ms$ star with the strongest attenuation curve from \citet{pei92}, which
is that of the Milky Way (MW).  The SMC and MW curves bracket the dust extinction that could 
realistically be expected in the afterglow spectra.  The star was placed in the largest core and 
cloud densities in our survey (10 and 10$^5$ cm$^{-3}$, respectively).  We find that the final radius 
of the H II region does not change with the new attenuation curve.  Since this choice of parameters is 
the most hostile to the formation of an H II region, we conclude that none of our other models would 
be affected over the range of extinction likely to be found in the galaxies.

\subsection{Winds}

In our models we have excluded the 500 - 1000 km s$^{-1}$ line-driven winds of the stars.  Winds would 
at least partially blow away the molecular envelope of the star, excavating a cavity of diffuse coronal 
gas at several million K and plowing a strongly shocked (and completely ionized) gas shell out into the 
cloud.  However, they would not alter the size of the H II regions at the time of the blast.  While it
is true that winds would disperse dense cores and facilitate breakout of UV into the cloud, there is 
little variation in final H II region diameters with $n_{\mathrm{c}}$ anyway.  Once the H II region encloses the 
bubble, it will continue to expand in spite of the dense shell driven by the wind because it is so hot 
it cannot recombine and attenuate the UV flux reaching the I-front.  For plausible cloud densities, the 
shell will not overtake the H II region (100 pc) prior to the burst.  Furthermore, dust would be 
evaporated inside the bubble but survive beyond it, with little difference to molecular fractions in the 
rest of the cloud.  Because the properties of the bubble are not observationally constrained by current 
afterglow data and their inclusion would change neither the minimum radii at which neutral H is observed 
nor the H$_2$ column densities in the sight lines, we do not include winds in this study.

\begin{figure}
\resizebox{3.45in}{!}{\includegraphics{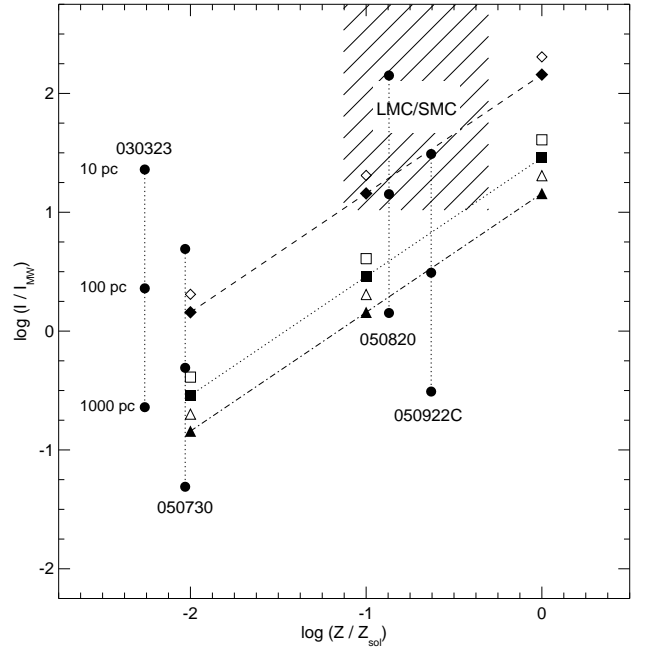}}
\caption{Minimum LW backgrounds required to suppress H$_2$ mass fractions below 10$^{-6.5}$ in our 
models, in units of the Milky Way average background.  Filled symbols:  ZEUS-MP; unfilled symbols: 
analytical.  Triangles: $n_{\mathrm{cl}}$ = 1 cm$^{-3}$; squares: $n_{\mathrm{cl}}$ = 5 cm$^{-3}$; 
diamonds: $n_{\mathrm{cl}}$ = 10 cm$^{-3}$; circles: the uniform LW backgrounds implied by eq 
\ref{eq:I_MW} for the afterglows in TET07, assuming diameters of 10, 100, and 1000 pc for 
the GRB host cloud.  The triple for each afterglow is connected by a vertical dotted line and is 
designated by its GRB.} 
\label{fig:fh2}
\end{figure}

Nevertheless, two important points about wind blown bubbles merit further investigation.  First, 
interactions between strong ionized flows in the H II region and the shell would lead to complex morphologies 
in the circumstellar environment of the GRB that could be crucial to its evolution \citep[\textit{e.g.} ][]{fet03}.  Future 
observations might unveil these geometries.  Second, UV and x-ray photons from the hot wind-blown shock could
modify the structure of the H II region \citep{ck07}.  The hard radiation broadens the I-front and creates a 
temperature front (T-front) precursor ahead of it \citep{qet07}.  The front widens because of the spread in 
photon mean free paths in the high-energy tail of the x-ray spectrum.  The T-front is the layer of hot, nearly 
neutral gas formed when the most energetic photons in the spectrum penetrate far into the gas upstream of the
front, ionize it and create secondary electrons capable of other collisional ionizations.  This minute but 
energetic flux can only sustain ionized fractions of 10$^{-3}$ - 10$^{-4}$ ahead of the front, but this is 
sufficient to raise their temperatures to 10$^4$ K, considerably hotter than the shock it precedes.  X-rays can 
either discourage or promote H$_2$ formation in the outer layer of the H II region.  The partial ionization and 
warm temperatures there are ideal for catalysis of H$_2$ via the H$^-$ channel \citep{rgs01}, but the high 
temperatures of the T-front may curtail its formation.

\section{Photodissociation of the Cloud by the Galactic Background}

We now explore the LW intensities necessary for H$_2$ destruction for a variety of column 
densities and metallicities that are consistent with observations of the GRB progenitor cloud.  
As noted in section \ref{sec:fuv}, we do not perform direct transport of galactic UV photons in
the outer regions of the cloud so we make no attempt to model the rise of either LW backgrounds or
H$_2$ fractions there.  Molecular hydrogen, dust, and dissociating photons all come to equilibrium 
in the cloud from a variety of initial abundances and intensities.  Since we ignore the evolution
of the background, there is some ambiguity to determining the minimum field necessary to drive H$_2$ below 
detection limits in the cloud.  Rather than address the many evolutionary possibilities for the 
background and whether it even arises given a set of initial circumstances, we proceed from the 
assumption that it does appear.  We then determine what intensities are required to suppress H$_2$ 
formation below detection thresholds for a particular neutral H column density and metallicity. This 
stance is logical, given that the absence of H$_2$ in the GRB sight lines renders the presence of LW 
radiation extremely likely and that its existence is of greater relevance than its origin, or history.  
However, note that in doing so we exclude some possibilities, such as large early H$_2$ abundances 
capable of quenching LW backgrounds.  
 
We adopt a cloud core density $n_{\mathrm{c}} of $ 10 cm$^{-3}$ and cloud densities $n_{\mathrm{cl}}$ 
of 5, 10, and 50 cm$^{-3}$.  We examine metallicities of 0.01, 0.1, and 1.0 $\Zs$, for a total of nine 
models.  Our coordinate grid was 200 uniform zones, with inner and outer boundaries of 0.1 and 1 kpc, 
respectively, and the same boundary conditions as in our ealier models.  Our choice of mesh accelerates 
convergence to the H$_2$ abundances at some cost in accuracy to the H II region, but this is of little
consequence to the molecular fractions in the cloud.  In each case the simulation is run until H$_2$ 
comes into equilibrium with the galactic background.  The H$_2$ mass fraction is initialized to 2.0 
$\times$ 10$^{-4}$.  Although not of direct importance to the molecular abundances, the time-dependent 
spectrum of the 40 $\Ms$ cluster was used in these runs.

The minimum background intensities required for H$_2$ suppression below the detection limit (defined 
to be f$_{H_2} = $10$^{-6.5}$) are plotted for all nine models in Fig \ref{fig:fh2}.  It is clear that 
thresholds rise with both ambient density and metallicity as expected, varying from 0.144 to 144 times 
the Galactic UV mean.  Initial H$_2$ fractions greater than 2.0 $\times$ 10$^{-4}$ merely cause the 
cloud to reach equilibrium later and do not alter the final abundances.  We have verified that our 
results hold with initial fractions as large as 0.2.  The threshold intensities for 0.1 $\Zs < Z < \Zs$
are consistent with the FUV backgrounds observed in star-forming regions in the LMC and SMC, which appear 
in the shaded region in Fig \ref{fig:fh2} \citep{tum02}. 

The scaling of threshold intensities with both cloud density and metallicity can be understood with 
simple equilibrium arguments.  The timescale for H$_2$ destruction in the cloud is the just the inverse  
of the photodissociation rate coefficient
\begin{equation}
t_{\mathrm{dis}} = {k_{\mathrm{dis}}}^{-1}, \vspace{0.1in}
\end{equation}
which is 1.84 $\times$ 10$^{10}$ s for the mean Milky Way UV background.  This is faster than any 
hydrodynamical process in the outer regions of the cloud, so final H$_2$ fractions are determined
by equilibrium between creation by dust (which completely dominates formation by the H$^-$ channel at 
these densities) and destruction by LW photons.  Taking maximum mass fractions f$_{H_2} = 10^{-6.5}$, 
one can approximate threshold intensities by equating the creation and destruction processes:
\begin{equation}
I_{\mathrm{MW}} = \displaystyle \frac{Z R \, n_{\mathrm{H}}}{7.77 \times 10^{5} f_{\mathrm{H}_2}}.
\label{eq:I_MW} 
\end{equation}
These equilibrium estimates, which we plot in Fig \ref{fig:fh2}, differ from the code values by at
most 30\,\%.  The intensities scale linearly with metallicity and density, as confirmed by ZEUS-MP.

A galactic LW background can be estimated for the afterglow spectra of TET07 using eq \ref{eq:I_MW}.  
The density $n_{\mathrm{H}}$ of the GRB host cloud can be fixed by 
assuming that the neutral H column density of the afterglow is local to the GRB and dividing it by 
some fiducial diameter for the cloud.  We chose 10, 100, and 1000 pc for the size of the cloud and 
adopted N$_{\mathrm{H}}$ from column 3 of Table 1 in TET07 for each spectrum.  The detection limit 
f$_{H_2}$ was computed from N$_{\mathrm{H}}$ and N$_{\mathrm{H_2,min}}$ from column 8 of Table 1.  
We plot the three intensity thresholds for H$_2$ destruction for 4 of the 5 GRBs in TET07 in Fig 
\ref{fig:fh2} as a function of metallicity (which are found in column 4 of Table 1).  

We find that the FUV backgrounds implied by eq \ref{eq:I_MW} for the afterglows are consistent with 
those in our models of comparable metallicity, particularly those associated with more realistic 
diameters of 10 - 100 pc for the GRB host cloud.  If we extrapolated our models to lower metallicity,
they would also include the LW backgrounds associated with GRB 030323 for somewhat larger but still
reasonable cloud sizes.  Perhaps of most interest, afterglows with metallicities comparable to those
of the SMC and LMC have similar uniform backgrounds, lending weight to the notion that GRBs originate
from active star-forming regions, even though their host galaxies cannot be observed. 
 
Our thresholds are valid for larger clouds at the same density due to the uniform nature of the 
background.  The converse of these results is also true: if the host cloud has LW backgrounds below 
these levels, the GRB afterglow will exhibit H$_2$ absorption regardless of the initial molecular 
fractions.  Since the borderline intensities are similar to those in regions of active star formation, 
at least in low metallicities, we anticipate it is only a matter of time before molecular hydrogen is 
detected in afterglows, if indeed GRB 060206 is not already an example of one \citep{fet06}. Since 
we neglect both transport and self-shielding of LW flux, which reduce the dissociation of H$_2$, our 
results somewhat underestimate the minimum backgrounds required to destroy molecular hydrogen.

\section{Conclusions}

The absence of H$_2$ in the large neutral H columns of GRB afterglow spectra is easily explained by the 
H II region of the progenitor or its cluster and by values of galactic LW backgrounds that are consistent
with star-forming regions.  In most cases the progenitor alone is sufficient to create the H II region.  
Since the H II region is illuminated when the progenitor dies, and because stars in the cluster form at 
nearly the same time, the minimum radius at which neutral H is observed in the sight lines is determined 
by both the lifetime of the progenitor and sound speed in the ionized gas rather than by the number 
of stars in the group or its composition.  As a result, these radii (as inferred from strong \ion{Mg}{1} 
absorption) as a rule will be 50 - 100 pc.  The sound speed (and therefore the expansion rate of the 
D-type front) is a relatively weak function of flux and spectral profile, so we expect these radii to 
be a robust feature of afterglow spectra.  The fairly narrow band in which they fall corresponds to the 
spread in MSL for likely GRB progenitors and is nearly identical for solitary stars and clusters, making 
it difficult to discriminate whether the star was in a group or alone.  Stars less massive than 25 $\Ms$ 
might fail to create an H II region, but they are not likely GRB candidates.  Our findings are consistent
with observed H II regions such as 30 Doradus in the LMC, whose radius is $\sim$ 150 pc and contains 
around 50 OB stars \citep{mh98}.

Total flux plays a role in the sizes of the H II regions only if it is so intense that the front is still 
R-type at the time of the burst.  Unrealistic numbers of massive stars would be necessary for this to happen, 
especially if dust is present.  Winds, although not included in these models, will reinforce our range of 
minimum radii by reducing their dependence on the density of the cloud core.  X-rays and extreme UV from the 
wind-blown shell may alter the structure of H II regions but not their radii or H$_2$ abundances in the  
cloud.

H$_2$ abundances in the ambient ISM are nearly independent of the properties of the GRB progenitor or host 
cluster.  They are set by the mean intensity of the UV background, dust abundance and gas density.  Since 
the minimum intensities required to extinguish H$_2$ are common to regions of vigorous star formation,
it is not surprising that molecular hydrogen has not yet been conclusively identified in GRB afterglows. 
On the other hand, these backgrounds are large enough that some clouds might not achieve them, so it is 
likely that H$_2$ will eventually be found along a GRB sight line. 

All of our models hold the GRB progenitor at the center of the molecular cloud core, but this in reality
is unlikely due to the peculiar motions of the stars in the cluster.  However, stellar motion within the 
cluster would be at most 10 pc and would still be bracketed by our models because the progenitor remains 
deeply imbedded within the cloud.  If the star wandered toward the edge of the core, the size of its H II 
region would depend only on $n_{\mathrm{cl}}$, not $n_{\mathrm{c}}$.  As our models show, the radius would 
not increase by more than 10\,\% in these circumstances.  The I-front would blow out the wall of the core, 
spilling ionized gas out into the cloud in a supersonic 'champagne' flow \citep[$\eg$][]{ytb84}.

One issue that remains unclear is the effect of the optical/UV afterglow itself on molecular hydrogen
in its path.  Although momentary, it is extremely intense (10$^{60}$ photons).  \citet{draine02} 
convincingly show that the optical flash in neutral molecular cores will destroy H$_2$ only out to 5 - 
10 pc and have little effect on molecular fractions in the rest of the host cloud.  However, this 
treatment does not account for the H II region, which hollows out a diffuse cavity of gas $\sim$ 100 pc
in radius.  What remains to be determined is if the optical pulse can destroy any H$_2$ beyond the H  
II region that was not dissociated prior to the burst.  If so, this would lower the minimum LW 
background necessary to extinguish molecular hydrogen in the rest of the cloud; if not, our results
would unambiguously prove that the H$_2$ was absent before the explosion.  This question cannot be 
settled by the current version of our code because the static approximation to radiative transfer we 
employ breaks down in the limit of these short, intense flashes.  Upgrades to our algorithm that 
retain the time-dependent term of the transfer equation will enable us to resolve these issues in
a future study.

Multidimensional simulations now in progress will reveal whatever dynamical instabilities occur in the 
ionization front and wind bubble.  Efficient radiative cooling by both metals and H$_2$ in the shocked 
neutral gas accumulated by the I-front will collapse it into a thin shell prone to fragmentation 
\citep{gsf96,wn07a}.  Rayleigh-Taylor instabilities may appear in the wind bubble even in the absence 
of cooling if it accelerates in density gradients \citep[\textit{e.g.} ][]{mn93}.  Both phenomena can 
clump gas on multiple spatial scales in the flow prior to the burst and might be imprinted on afterglow 
spectra.  Future observations may uncover these features and yield insights into the environment of the 
burst.

Finally,  it will be valuable to compare the incidence of H$_2$ detections in 
GRB host galaxies with the incidence measured in QSO  
damped Lyman-alpha (DLA) sight lines. The latter show an incidence
of $\approx 10\%$ for molecular fractions $f({\rm H_2}) > 10^{-4.5}$ \citep{nlp+08}.   
Current expectation is that the QSO-DLA sightlines penetrate 
galaxies that have lower total SFRs at a larger average impact parameter
\citep{pcd+07,ftps+08}.  
The results presented here suggest that one may test these hypotheses, in part, 
by measuring the relative incidence of H$_2$. 
For example, one may expect a lower average UV intensity along
QSO-DLA sightlines than GRB-DLA sightlines and therefore a higher
incidence of H$_2$ detections.

\acknowledgments

DW thanks Hsiao-Wen Chen, Mark Krumholz, and Michael Norman for helpful discussions concerning these 
simulations.  This work was carried out under the auspices of the National Nuclear Security Administration of 
the U.S. Department of Energy at Los Alamos National Laboratory under Contract No. DE-AC52-06NA25396.  
J. X. P. is partially supported by NASA/Swift grants NNG06GJ07G and NNX07AE94G and an NSF CAREER 
grant (AST-0548180).  AH was supported by the DOE Program for Scientific Discovery through Advanced 
Computing (SciDAC; DE-FC02-01ER41176).  JT gratefully acknowledges the support of Gilbert and Jaylee 
Mead for their namesake fellowship in the Yale Center for Astronomy and Astrophysics.  The simulations 
were performed on the open cluster Coyote at Los Alamos National Laboratory.

\end{document}